\documentclass[5p,twocolumn]{elsarticle}
\usepackage{graphics,graphicx}
\usepackage{mathrsfs}
\usepackage{amsmath,bm,amsfonts,amssymb,wasysym}
\usepackage{color}
\usepackage{natbib}
\usepackage{epstopdf}
\usepackage{epsfig}
\usepackage{bm}
\usepackage{listings}
\usepackage{graphicx}
\usepackage{newtxtext}
\usepackage{newtxmath}
\usepackage{hyperref}
\usepackage{subcaption}
\usepackage{comment}
\usepackage{cuted}

\biboptions{longnamesfirst,sort&compress}

\hypersetup{
    colorlinks = true,
    urlcolor   = blue,
    citecolor  = black, 
}

\newcommand{\RomanNumeralCaps}[1]
\linenumbers

\def\half{\mbox{$1\over2$}}
\def\quar{\mbox{$1\over4$}}

\DeclareMathOperator{\mF}{\mathcal{F}}
\DeclareMathOperator{\mW}{\mathrm{De}}
\DeclareMathOperator{\mR}{\mathcal{R}}

\DeclareMathOperator{\mL}{\mathcal{L}}

\DeclareMathOperator{\mJ}{\mathcal{J}}
\DeclareMathOperator{\Df}{\mathcal{D}}
\DeclareMathOperator{\bet}{\alpha}

\def\R0{{R_0}} 
\def\LL{{L}} 

\def\thir{{\mbox{$1\over3$}}}

\newcommand{\pd}[2]{\frac{\partial #1}{\partial #2}}

\newcommand{\be}{\begin{equation}}
\newcommand{\ee}{\end{equation}}

\journal{Journal of Non-Newtonian Fluid Mechanics}

\begin{document}

\begin{frontmatter}

\title{Rayleigh-Plateau instability of an elasto-viscoplastic filament}

\author[]{J. D. Shemilt\corref{cor1}} 
\author[]{N. J. Balmforth} 

\address{ Department of Mathematics, University of British Columbia,
  Vancouver, BC, V6T 1Z2, Canada}
\cortext[cor1]{Corresponding author: {\it E-mail:} shemilt@math.ubc.ca}

\date{\today}

\begin{abstract}
  A slender-thread model is used to explore the Rayleigh-Plateau
  instability of a filament of elasto-viscoplastic fluid.
  Without elasticity, a finite yield stress suppresses any linear instability
  for a filament of constant radius. Including sub-yield elastic
  deformation permits an elastic Rayleigh-Plateau instability
  above a critical Deborah number. 
  If stresses over the thinner sections
  of the thread breach the yield threshold, viscoplastic deformations
  then drive the filament towards pinch-off.
  The thread consequently evolves to a beads-on-a-string structure.
  The elasto-plastic anatomy of the beads is explored and categorized.
\end{abstract}

\end{frontmatter}

\numberwithin{equation}{section}

\section{Introduction}

In the classical Rayleigh-Plateau instability, a circular thread
of viscous fluid becomes unstable towards varicose perturbations which
amplify until the thread pinches off over the narrowest sections
\cite{eggers08}.
For very viscous threads, pinch-off arises in finite time
\cite{eggers93,papageorgiou95}.
If the thread is made of a visco-elastic liquid,
the Rayleigh-Plateau instability again appears, although
thinning generates strong extensional stresses that oppose
pinch-off ({\it e.g.} \cite{goldin69,renardy95,entov84,li03,clasen06,turkoz18}).
This leads
to the formation of a ``beads-on-a-string''
structure, in which most of the fluid becomes concentrated into
nearly spherical beads connected by strings that thin exponentially
with time, unless nonlinear elastic effects are taken into account
\cite{renardy95,fontelos04}.

For fluids with a yield stress, the situation can be quite different:
if deformations below the yield stress are discarded, an initially
uniform thread lies relatively far from the yield threshold
because capillary forces are countered purely by elevated internal pressures,
not deviatoric stresses.
Arbitrarily small varicose perturbations cannot then
induce sufficient stress to breach the yield threshold. Consequently,
the Rayleigh-Plateau
instability becomes completely suppressed \cite{dripsI},
in a similar manner to how a number of different hydrodynamic instabilities
become eliminated by yield stresses in other settings \cite{annrev,shemilt2022surface,shemilt2023surfactant}.

That said, any real material does deform below the yield stress,
most typically visco-elastically. Moreover, 
surface tension can drive Rayleigh-Plateau-type instabilities
in threads of soft solids
\cite{matsuo92,barriere,mora2010capillarity,taffetani15a,fu21,pandey21}.
Therefore, it remains possible for this instability to re-assert
itself if viscoelastic deformations below the yield stress
are taken into account. The goal of the current article
is therefore to conduct a study of the
Rayleigh-Plateau instability in a thread of elasto-viscoplastic fluid.
\textcolor{black}{Applications include the atomization and printing 
  of such fluids \cite{goldin72,sauret26}.
  The instability can also play a role in other elasto-viscoplastic flows
  in which thin fluid jets are generated, such as bubble bursting
  \cite{sanjay21,balasubramanian24} or droplet impact \cite{peng25}. }

\textcolor{black}{To describe the elasto-viscoplastic rheology of our threads,} we appeal to the family of constitutive models
proposed by Saramito \cite{saramito07,saramito09}.
These models combine either the Bingham or Herschel-Bulkley law
from viscoplasticity \cite{annrev} with the Oldroyd-B model
from viscoelasticity. Saramito's model contains a switch
equivalent to the von Mises yield condition. Below
the implied stress threshold, the material behaves like
a Kelvin-Voight elastic solid; above the yield point, the fluid
behaves elasto-viscoplastically. The former behaviour allows
for linear Rayleigh-Plateau instability to act as a trigger
for the pinch-off of a thread.

In more detail and following common practice
in the interrogation of the fluid mechanics
of slender filaments \cite{eggers08}, we base our analysis
on a slender-thread model. Thus, we first outline a simplification
of the full governing equations of the
elasto-viscoplastic thread that furnishes a reduced model
that is more compact and straightforward to analyse (\S \ref{sec:model}).
The reduction follows a standard pathway of asymptotics, and
parallels the reduction for thin viscoplastic threads \cite{dripsI}.
The model itself was previously stated by Zakeri {\it et al.}
\cite{zakeri25}, although they did not
consider the Rayleigh-Plateau instability or
provide any detailed solutions describing the full dynamics
of a pinching thread (see also \cite{moschopoulos23}).
Here, we accomplish those tasks in \S \ref{sec:linstab}
and \S \ref{sec:nonlin}.
More specifically, in \S \ref{sec:linstab}
we connect the dynamics of the linear
instability captured by the model with the
elastic version of Raleigh-Plateau 
\cite{barriere,mora2010capillarity},
then explore the nonlinear dynamics that can lead to pinch-off
in \S \ref{sec:nonlin}. As long as yielding takes place
in the thread, that filament evolves towards a beads-on-a-string
structure whose anatomy we interrogate more thoroughly
in \S \ref{sec:anato}. Finally, we provide a fuller
description of the pinching dynamics in \S \ref{sec:carpetbomb}, \S\ref{solv} \textcolor{black}{ and \S\ref{sec:long}},
where we avoid specific choices of the problem parameters
and explore the model more generally.


\section{Slender-thread model}\label{sec:model}

\subsection{Governing equations}

\begin{figure}
    \centering    \includegraphics[width=\linewidth]{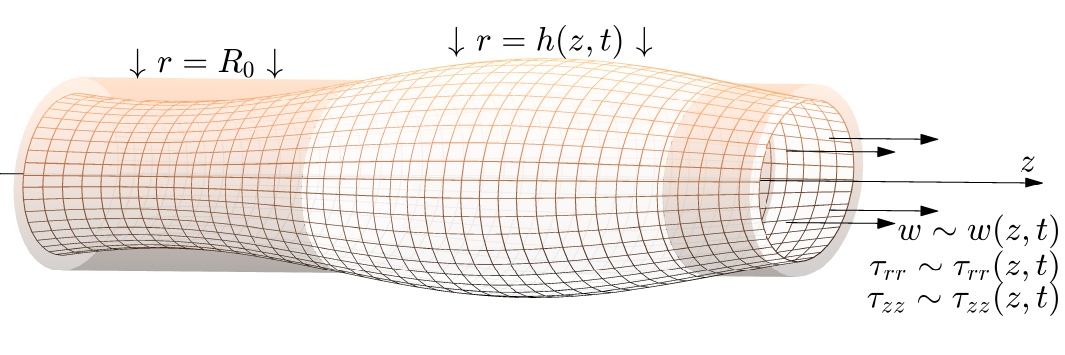}
    \caption{Sketch of the geometry of a thin thread, showing the
       main model variables.}
    \label{fig:geomsketch}
\end{figure}

Consider an axisymmetric thread of elasto-viscoplastic fluid
described by cylindrical coordinates $(r,\theta,z)$, as sketched in
Fig.~\ref{fig:geomsketch}. We
denote the velocity, pressure, and the total stress tensor within the thread
by $(u,0,w)$, $p$, and $\boldsymbol\sigma-p\boldsymbol{I}$, respectively.
In the absence of any body forces,
conservation of mass and momentum demand that
\begin{eqnarray}
 0 &=& \frac{1}{r}\frac{\partial}{\partial r}(ru) + \frac{\partial w}{\partial z},\label{masscons}\\
 \rho\frac{D u}{Dt}&=& -\frac{\partial p }{\partial r}
 + \frac{1}{r}\frac{\partial}{\partial r}(r\sigma_{rr}) - 
 \frac{1}{r}\sigma_{\theta\theta} + \frac{\partial}{\partial z}\sigma_{rz},
 \label{rmom1}\\
 \rho\frac{Dw}{Dt} &=&  -\frac{\partial p }{\partial z} +
 \frac{1}{r}\frac{\partial}{\partial r}(r\sigma_{rz}) +
 \frac{\partial}{\partial z}\sigma_{zz}.\label{zmom1}
\end{eqnarray}

We adopt Saramito's constitutive law \cite{saramito07}
for the rheology of the thread, defining
\begin{equation}
  \boldsymbol\sigma 
  = \boldsymbol\tau 
  + \mu_s \dot{\boldsymbol{\gamma}} 
  ,\label{sigma}
\end{equation}
where $\tau_{ij}$ is a polymeric component of the stress, and $\mu_s \dot{\boldsymbol{\gamma}}_{ij}$ is a solvent viscous component with
solvent viscosity $\mu_s$. Here,
\be
\dot{\boldsymbol{\gamma}}  
= \begin{pmatrix}
  2\frac{\partial u}{\partial r} & 0 &
  \frac{\partial w}{\partial r}+\frac{\partial u}{\partial z} \cr
  0 & 2\frac{u}{r} & 0 \cr
  \frac{\partial w}{\partial r}+\frac{\partial u}{\partial z} & 0 &
  2\frac{\partial w}{\partial z}
\end{pmatrix}
.
\ee
 The polymeric stress obeys
 \begin{equation}
  \lambda\overset{\triangledown}{\boldsymbol{\tau}} + 
                K {\boldsymbol{\tau}}
  = \mu_p\dot{{\boldsymbol{\gamma}}},\qquad
  K \equiv \max\left(\frac{\tau_d - \tau_Y}{\tau_d},0\right)
  ,
 \label{taupeqn1}  
\end{equation}
where $\overset{\triangledown}{\boldsymbol{\tau}}$
denotes the upper convected derivative,
$\lambda$ is a relaxation time, $\tau_\mathrm{y}$ is the yield stress,
$\mu_p$ is a plastic viscosity and
\be
\tau_d = \sqrt{\half\sum_{i,j}
  \left[\tau_{ij} - \thir{\rm Tr}({\boldsymbol{\tau}})\right]^2} 
\ee
is the second invariant of the deviatoric part of $\boldsymbol{\tau}$.
Note that, when $\lambda=0$, \eqref{taupeqn1} reduces to the Bingham model,
with constant plastic viscosity; the generalization
proposed in \cite{saramito09} allows for a power-law-type viscosity, but we
used the simpler version here.

In line with standard discussions of the Rayleigh-Plateau instability,
we consider a spatially periodic thread;
the initial radius is $\R0$ and each periodic segment has length $\LL$.
We denote the surface of the thread by $r=h(z,t)$ and ignore any air
resistance. Here, the kinematic condition demands that
\be
\pd{h}{t} + w \pd{h}{z} = u \label{kinbc}
.
\ee
We quote the stress boundary conditions more explicitly later, but
in a simplified form suited to the slender limit.

\subsection{Slender-thread theory}

For a slender thread, the aspect ratio is small:
$\epsilon = \R0/\LL\ll1$. In this limit,
the radial velocity $u$ must be $O(\epsilon)$
smaller than the axial velocity $w$.
Moreover, because shear
stresses cannot be effectively developed across the thread, the axial
flow is plug-like with $w=w(z,t)+O(\epsilon^2)$.
Mass conservation \eqref{masscons} then requires that
\begin{equation}
  u = -\half r\frac{\partial w }{\partial z} ,
  \label{urel}
\end{equation}
to leading order.
Given these scalings, the
components, $\{\dot{\gamma}_{rr}, \dot{\gamma}_{\theta\theta}, \dot{\gamma}_{zz}\}$
dominate the tensor $\dot{\boldsymbol{\gamma}}$, 
whilst $\dot{\gamma}_{rz}$ is $O(\epsilon)$ smaller. The polymeric shear stress
$\tau_{rz}$ is therefore unforced at leading order
in \eqref{taupeqn1}, so there is a consistent solution in which
that shear stress remains $O(\epsilon)$ smaller than the normal
stresses $\{\tau_{rr}, \tau_{\theta\theta}, \tau_{zz}\}$.
Similarly, those components can be taken to be independent of $r$
to leading order.
The evolution equations and boundary conditions
for $\tau_{rr}$ and $\tau_{\theta\theta}$
now become identical, given \eqref{urel}, and
so if $\tau_{rr}=\tau_{\theta\theta}$ initially, those components remain equal
at later times. We focus on initial-value problems in which all components
of the polymer stress vanish, and therefore
take $\tau_{rr}=\tau_{\theta\theta}$. From \eqref{urel}, $\dot{\gamma}_{rr} = \dot{\gamma}_{\theta\theta}$ to leading order, so we also have $\sigma_{rr}=\sigma_{\theta\theta}$. 

At leading order, \eqref{rmom1} and \eqref{zmom1} then imply 
\begin{eqnarray}
     0 &=& - \frac{\partial p}{\partial r} + \frac{\partial\sigma_{rr}}{\partial r},\label{rmom2}\\
     \rho \left(\frac{\partial w}{\partial t} + w\frac{\partial w}{\partial z}\right)&=& - \frac{\partial p}{\partial z} + \frac{\partial\sigma_{zz}}{\partial z} + \frac{1}{r}\frac{\partial}{\partial r}(r\sigma_{rz})
     .\label{zmom2}
\end{eqnarray}
The stress evolution equation \eqref{taupeqn1}
now boils down to the two independent relations,
\be
\begin{aligned}
  \lambda\left(\frac{\partial\tau_{rr}}{\partial t}
  + w\frac{\partial\tau_{rr}}{\partial z} - 2\frac{u}{r}\tau_{rr}\right)
  + Y\tau_{rr} &= -\mu_p\frac{\partial w}{\partial z},
  \\
  \lambda\left(\frac{\partial\tau_{zz}}{\partial t}
  + w\frac{\partial\tau_{zz}}{\partial z}
  - 2\frac{\partial w}{\partial z}\tau_{zz}\right)
  + Y\tau_{zz} &= 2\mu_p\frac{\partial w}{\partial z},
\end{aligned}
\label{constA}
\ee
in which we may now take
\begin{equation}
  Y = \max\left(\frac{|\tau_{rr}-\tau_{zz}|-\sqrt{3}\tau_Y}{|\tau_{rr}-\tau_{zz}|}
  ,0\right).
\end{equation}

The leading-order stress conditions at the free surface are
\be
\left.
\begin{aligned}
    p - \sigma_{rr}-\Sigma\kappa &=&0\\
    \sigma_{rz} + \frac{\partial h}{\partial z}(p-\sigma_{zz}-\Sigma\kappa) &=& 0
\end{aligned}
\right\} \quad {\rm on}\ r=h,
\label{stressBC}
\ee
where the surface curvature is given by
\begin{equation}
  \kappa = \frac{1}{h[1+(\partial_zh)^2]^{1/2}}
  - \frac{\partial_{zz}h}{[1+(\partial_zh)^2]^{3/2}}.
  \label{kappa}
\end{equation}
and $\Sigma$ is the surface tension. We retain the full expression for the curvature \eqref{kappa} (not just the leading-order term, $h^{-1}$)
following common practice in long-wave theory
({\it e.g.} \cite{eggers1994drop}).

It follows from \eqref{rmom2} and \eqref{stressBC} that
$p = \sigma_{rr}+\Sigma\kappa$ throughout the thread. The pressure
can then be eliminated from \eqref{zmom2}, and that relation
integrated over the cross-section of the thread to
furnish an evolution equation for the leading-order
axial velocity $w\sim w(z,t)$.
In combination with \eqref{constA}, \eqref{masscons} and \eqref{kinbc}, we
then arrive at the slender-thread model. We summarize this model
below, after placing it into a dimensionless form.


\subsection{Dimensionless model}

We define dimensionless variables,
\be
\begin{aligned}
  \hat{t} = \frac{t}{T}, &\quad (\hat{z},\hat{h}) = \R0^{-1}(z,h),\quad
  \hat{w} = \frac{T}{\R0}w, \\ &
  (\hat{p},\boldsymbol{\hat{\tau}}) = \frac{T}{\mu_p}({p},\boldsymbol{{\tau}}),
  \quad \hat{\kappa} = \R0\kappa,
\end{aligned}
\label{nondim}
\ee
where $T=\mu_p\R0/\Sigma$ is a capillary time scale.
After dropping the hat decorations,
we arrive at the dimensionless model equations,
\be
\begin{aligned}
  h_t + wh_z &= - \half hw_z, 
  \cr
  \mR\left(w_t + ww_z\right) &= \frac{1}{h^2}
          [h^2(\mF\{h\}-\tau_{rr}+\tau_{zz}+3\alpha w_z)]_z 
          , 
  \cr
  \mW(\tau_{rr,t} + w & \tau_{rr,z} + w_z\tau_{rr}) + Y\tau_{rr} = - w_z,
  \\
  \mW(\tau_{zz,t} + w & \tau_{zz,z} - 2w_z\tau_{zz}) + Y\tau_{zz} = 2w_z,
\end{aligned}
\label{modeqs}
\ee
where
\begin{equation}
    Y \equiv \max\left(\frac{|\tau_{rr}-\tau_{zz}| - \sqrt{3}\mJ}{|\tau_{rr}-\tau_{zz}|},0\right),\label{Kdefn}
\end{equation}
the gradient of the curvature has been written as
$$
\kappa_z = -h^{-2} (h^2\mF\{h\})_z,
$$
with
\begin{equation}
  \mF\{h\} = \frac{h_{zz}}{[1+(h_z)^2]^{3/2}} + \frac{1}{h[1+(h_z)^2]^{1/2}},
  \label{Frel}
\end{equation}
and we have used subscripts of $z$ and $t$ as a shorthand notation
for partial derivatives, except where they identify a stress component.
The dimensionless groups that appear in \eqref{modeqs}-\eqref{Kdefn} are
defined by
\begin{equation}
     \mW = \frac{\lambda\Sigma}{\mu_p\R0},\quad \mJ = \frac{\tau_Y\R0}{\Sigma},\quad \bet = \frac{\mu_s}{\mu_p}, \quad \mR = \frac{\rho \R0 \Sigma}{\mu_p^2},
\end{equation}
and are equivalent to
the Deborah number, plastocapillarity number, viscosity ratio and
inverse square of the Ohnesorge number, respectively.
We denote the dimensionless length of the thread by $\mL=\LL/\R0$.
Up to some differences in notation, the slender-thread model in
\eqref{modeqs}-\eqref{Kdefn} was stated previously by
\cite{zakeri25}.

In practice, we solve the model equations numerically, using
centred differences on a uniform spatial grid to replace axial
derivatives. The resulting system of ordinary differential equations
is then advanced in time using a standard stiff integrator (Matlab's
ODE15s). To ease the integration over longer times, we add
  diffusion terms to the stress equations. However, we further verified
  that this addition did not significantly affect the solutions;
  see \ref{app:numerics}. 
By way of initial condition, we set
\be
\begin{aligned}
h(z,0)&=1+10^{-3}\cos(2\pi z/\mL), \\
w(z,0)&=\tau_{rr}(z,0)=\tau_{zz}(z,0)=0,
\end{aligned}
\label{ICs}
\ee
in order to explore the evolution of the thread from a low-amplitude
initial state.

The initial conditions \eqref{ICs} are symmetric about the points $z=\{0,\mL/2\}$, and the model equations do not introduce any asymmetry, so the solution must remain symmetric about these points for all time. Therefore,
to reduce computation times for a given spatial resolution (we typically use $4000$ grid points in the computational domain),
we solve the model equations on the half-domain, $0\leq z\leq\mL/2$, subject to symmetry boundary conditions at the lateral ends,
  \be
\begin{aligned}
	h_z(0,t) &= h_z(\half\mL,t) = w(0,t) = w(\half\mL,t) = 0,\\
	\tau_{rr,z}(0,t) &= \tau_{rr,z}(\half\mL,t) = \tau_{zz,z}(0,t) = \tau_{zz,z}(\half\mL,t) = 0.
	\label{latbcs}
\end{aligned}
\ee
We terminate the computations if $\min(h)=0.02$, which is small enough to ensure that the thread has entered its late-time dynamical regime and evolution
elsewhere has mostly subsided. When the minimum radius does not approach zero at late times, instead relaxing towards a non-zero value, we stop computations if $\min(h)$ changes by less than $5\times10^{-6}$ during one unit of time. 

Altogether the model contains the five parameters: $\{\mW,\mJ,\mL,\mR,\bet\}$.
Our interest here is mostly on the impact of the first two,
which control the strength of elasticity and the yield stress.
Therefore, we mostly pick representative values
for the last three parameters, $\{\mL,\mR,\bet\}=\{20,0.01,1\}$,
and vary $\mW$ and $\mJ$. As we see below, the
choice for $\mL$ allows for threads that
are sufficiently long for Rayleigh-Plateau instability to take hold
unless elasticity is sufficiently weak ($\mW$ low enough)
and a yield stress is present ($\mJ>0$). 
The selection for $\mR$ is guided by our interest in problems
with relatively low inertia, whilst keeping the terms
on the left of (\ref{modeqs}b) to ease numerical computations. The choice of $\bet=1$ has no particular significance, but sets the solvent viscosity equal to the polymer viscosity. We discuss the effect of varying $\bet$ in \S\ref{solv}, \textcolor{black}{and of varying $\mL$ in \S\ref{sec:long}}. 


\section{Linear stability}\label{sec:linstab}

Consider small perturbations to a
uniform thread with
\begin{equation}
  (h,w,\tau_{rr},\tau_{zz}) =
  (1,0,0,0)+
(\check{h},\check{w},\check{\tau}_{rr},\check{\tau}_{zz})  e^{ikz+st},
  \label{LSAhpert}
\end{equation}
where $s$ is the growth rate and $k=2\pi/\mL$ is the wavenumber.
Introducing this decomposition into \eqref{modeqs} and
then linearizing in the amplitudes
$(\check{h},\check{w},\check{\tau}_{rr},\check{\tau}_{zz})$
yields the dispersion relation,
\begin{equation}
  \mR s^2 + 3k^2\bet s + 3k^2\left(\frac{1}{\mW} - \frac{1-k^2}{6}\right) = 0
  .
  \label{dispo}
\end{equation}
Note that the linearization immediately requires us to set $Y=0$, since an infinitesimal perturbation is not sufficient to yield the fluid for any finite $\mJ>0$, and therefore we are placed in the visco-elastic regime below the yield condition. As long as there is a non-zero yield stress, $\mJ$ has no impact on this linear stability analysis. 

\begin{figure}
    \centering    \includegraphics[width=\linewidth]{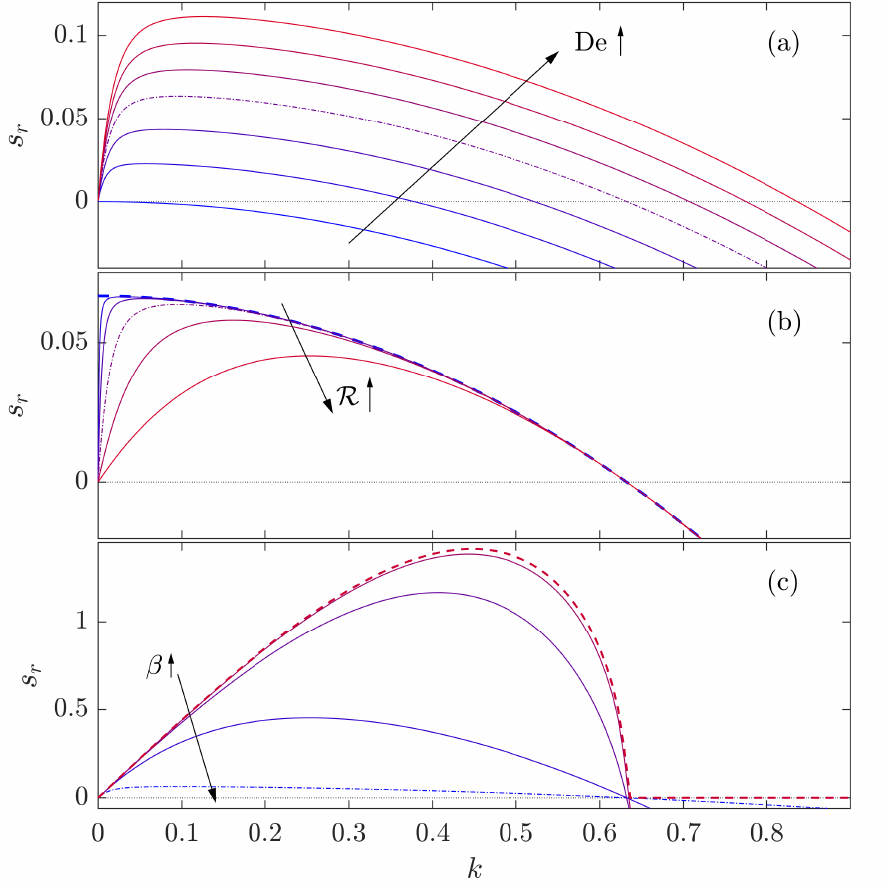}
    \caption{\textcolor{black}{Linear growth rates $s_r=\Re(s)$
        from \eqref{dispo} for (a) $(\mR,\bet)=(0.01,1)$ and $\mW=\{6,7,8.25,10,12,15,20\}$, (b) $(\mW,\bet)=(10,1)$ and $\mR=\{10^{-4},10^{-3},0.01,0.1,1\}$, and (c) $(\mW,\mR)=(10,0.01)$ and $\bet=\{10^{-3},0.01,0.1,1\}$. In each panel, the dot-dashed line corresponds to the parameter values used in figures \ref{fig:egb}-\ref{fig:shapej}. For (b), the dashed line shows the 
        growth rate for no inertia ($\mR=0$), and that in (c) for
    no solvent viscosity ($\bet=0$).} }
    \label{fig:dispr}
\end{figure}

\begin{figure*}[ht]
    \centering    \includegraphics[width=\linewidth]{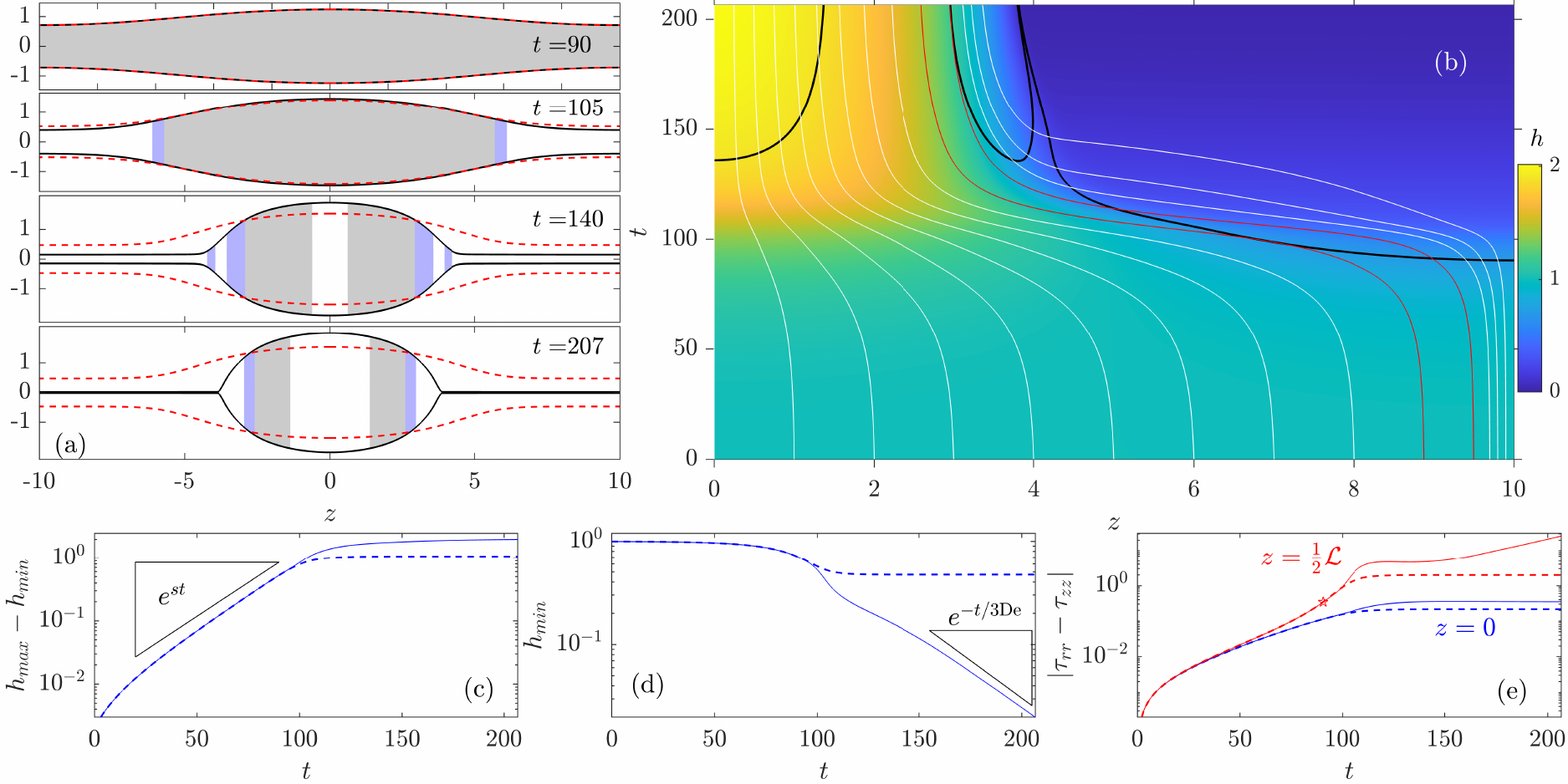}
    \caption{Sample solution for
      $\{\mW,\mL,\mJ,\bet,\mR\}=\{10,20,0.2,1,\textcolor{black}{0.01}\}$.
      (a) Snapshots of the thread at the times indicated. The elastic regions
      are shaded grey when never previously yielded, and blue otherwise. \textcolor{black}{Yielded regions within the thread are shown in white. Red dashed lines show the profiles for a thread that never yields ($\mJ\gg1$). Supplementary video 1 also shows the thread's evolution. } 
      (b) Radius plotted as a density on a space-time diagram,
      with superposed contours (black) showing the yield
      surfaces. The white lines indicate the paths taken by a selection
      of fluid elements; the red lines indicate the two fluid elements
      that border the region that passes through the yielded
      thinned section of the thread, but then falls below the
      yield stress and becomes elastic for later times.
      (c,d,e) Time series of $h_{max}-h_{min}$, $h_{min}$ and
      $|\tau_{rr}-\tau_{zz}|$ at $z=0$ (blue) and $z=\frac12\mL$ (red). 
      In (c), the triangle shows the expected linear growth rate;
      that in (d) shows the exponential decay $e^{-t/(3\mW)}$
      expected over the thinnest part of the thread.
      The star in (d) identifies the moment that the stress over thinnest
      part of the thread breaches the yield stress.
      The dashed lines in (d-e) show the corresponding
      time series for a thread that never yields ($\mJ\gg1$).
}
    \label{fig:egb}
\end{figure*}

Applying Descartes' rule of signs to
the dispersion relation \eqref{dispo} implies that there is one positive
real root when the final term is negative (all complex roots have
negative real part). Thus, the thread is
linearly unstable if
\begin{equation}
    \mW > \mW_c = \frac{6}{1-k^2}. \label{LSAcritW}
\end{equation}
To leading order in the long-wave limit
($k\to0$), the threshold in \eqref{LSAcritW} is that same as that
derived by Mora et al. \cite{mora2010capillarity} for the linear instability
of an elastic cylinder towards perturbations with arbitrary aspect
ratio. Their threshold actually reduces to
$\mW > (6+ k^2)/(1-k^2)+O(k^4)$, with the discrepancy in the
$O(k^2)$ terms originating from our long-wave approximation.

\textcolor{black}{The dispersion relation \eqref{dispo} is illustrated for a range of parameter values in figure \ref{fig:dispr}. As long as $\mW>\mW_c$, the growth rate is maximized at a finite value of $k$ that depends on the parameters $(\mW,\mR,\bet)$.
The most unstable wavenumber is $O(\mR^{1/2})$ for $\mR\ll1$.
When computing nonlinear solutions to the evolution equations in the
following sections, we choose, for simplicity, a domain length of
$\mL = 20$. The minimum wavenumber is then $k = \pi/10$, which
does not therefore correspond to the most unstable linear
perturbation for all our parameter settings, although it is
largely comparable.}


If $\mJ=0$ and the fluid has no yield stress, the linear stability
character outined above changes somewhat: because we must now take
$Y=1$ in the constitutive equations (\ref{modeqs}c,d), the
dispersion relation features an addition term $3k^2/[\mW(s+1)]$
on the right-hand side of \eqref{dispo}. The condition for linear instability 
in \eqref{LSAcritW} then becomes replaced by the simpler condition
$1-k^2 > 0$. That is, instability arises for $2\pi < \mL$, or
provided only that the thread is sufficiently long, as in the Newtonian
version of the problem ({\it cf.} \cite{goldin69}). \textcolor{black}{The maximum growth rate is also always attained at $k = 1/\sqrt{2}$ for
$\mJ=\mR=0$, unlike the $\mJ>0$ case discussed above, which requires
  inertia for the most unstable mode to have finite wavelength
  (see figure \ref{fig:dispr}b).}

\section{Nonlinear evolution}\label{sec:nonlin}

A sample numerical solution to the slender-thread model is shown in
Fig.~\ref{fig:egb}. In this example, the threshold \eqref{LSAcritW}
is met, so that the thread is linearly unstable to an
elastic Rayleigh-Plateau instability. This instability grows
exponentially with a rate predicted by the positive
solution to \eqref{dispo} (see Fig.~\ref{fig:egb}(c))
until the stress difference $|\tau_{rr}-\tau_{zz}|$ breaches the
yielding threshold over the thinnest sections of the thread
(Fig.~\ref{fig:egb}(b,e)).
Those sections then begin to thin more dramatically, with 
$|\tau_{rr}-\tau_{zz}|$ continuing to increase there
(Fig.~\ref{fig:egb}(d,e)). Eventually, some yielding also occurs over the
thickened sections of the thread.
By the end of the computation, most of the fluid
has collected into a periodic array of beads connected
by thin strings, in a similar fashion to that seen for a viscoelastic
thread \cite{clasen06,eggers20}. The beads evolve to largely steady
states, fed increasingly less by the waning flux from the strings
(Fig.~\ref{fig:egb}(c,e)).

Because axial normal stresses continually increase over the strings, the yield
stress eventually becomes irrelevant there
({\it cf.} \cite{dripsI,zakeri25}).
Consequently,
as for a thread of Oldroyd-B fluid \cite{renardy95,clasen06},
the strings thin exponentially at late times
with a dependence of $e^{-t/(3\mW)}$,
as seen in Fig.~\ref{fig:egb}(d).
This thinning contrasts sharply with the fate of the Rayleigh-Plateau
instability in soft elastic solids, for which both the
beads and strings evolve to steady states with finite radii
\cite{taffetani15a,pandey21}.
Indeed, Fig.~\ref{fig:egb} also includes results for a thread
in which $\mJ\gg1$. In that case, the fluid never yields,
the strings stop thinning 
and a steady state is reached everywhere over longer times.
Evidently, the yielding of the thinned sections provides a pathway
to the pinch-off of an elasto-viscoplastic thread
due to Rayleigh-Plateau instability,
albeit after an exponentially long time for the
current model. Zakeri {\it et al.} \cite{zakeri25} point out
how the inclusion of a power-law plastic viscosity can prompt
pinch-off in finite time.

For a visco-elastic thread, the beads eventually converge towards
spherical drops \cite{clasen06}. Here, this is not the case. Indeed, the
structure of the bead remains rather more complicated, reflecting
finer details of the elasto-plastic balances arising
at late times. In the sample solution shown in 
Fig.~\ref{fig:egb}, the bead initially remains mostly elastic
(see the snapshot at $t=105$ in panel (a)). 
However, yielded regions then migrate into the bead from the connections
to the adjacent strings, and the curvature of the core of the bead
becomes sufficient to force fluid to also yield there
({\it cf.} the snapshots at $t=140$ and 207). For long times,
the stresses over the yielded sections decline back to
the yield point, leaving plastic relics that sandwich
the remaining elastic parts of the bead.

\begin{figure*}
    \centering    \includegraphics[width=\linewidth]{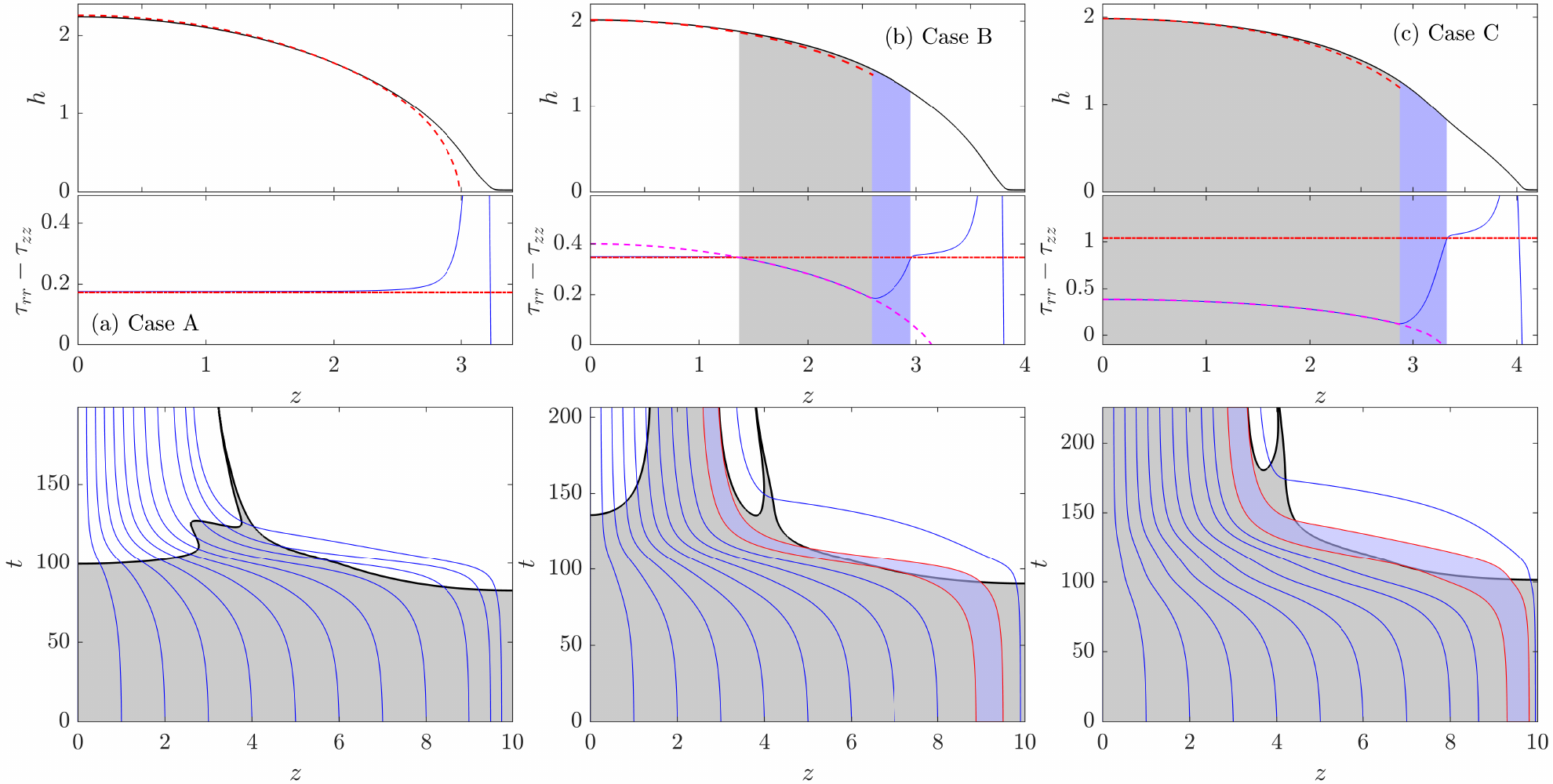}
    \caption{Numerical solutions showing examples of
      (a) case A ($\mJ=0.1$), (b) case B ($\mJ=0.2$) and
      (c) case C ($\mJ=0.6$), all with
      $\{\mW,\mL,\bet,\mR\}=\{10,20,1,0.01\}$.
      For each case, the upper rows of panels show the
      final shapes and stress differences, whilst the bottom
      row presents the yield surfaces (thicker black lines)
      and the trajectories of sample fluid elements (thinner blue lines)
      on space-time diagrams. In the upper rows, the red dashed
      lines show solutions of \eqref{eq:finprof}, the magenta dashed lines in (b,c) show the elastic stress
      state in \eqref{elastotaudif}, and the (red) dot-dashed
      lines indicate the yield condition $\tau_{rr}-\tau_{zz}=\sqrt3 \mJ$.
      The shaded regions correspond to the elastic regions,
      either in grey if never yielded, or blue if previously so.
      In the bottom row, the red lines bordering the blue-shaded area
      show the fluid elements buffering the region that yields {\it en
        route} to a final elastic state.
    }
    \label{fig:abc}
\end{figure*}

The elasto-plastic control of the final shape depends
sensitively on the deformation history of the bead, as illustrated
in Fig.~\ref{fig:abc}. This figure compares the sample
solution shown earlier with two further examples, and displays three
possible outcomes of the pinching process. The earlier
example, denoted as case B, ends with a partly elastic bead containing
a yielded core and borders.
With a smaller yield stress, as in case A, the entire bead
becomes yielded, leaving an entirely plastic relic at long times.
For stronger yield stress, as illustrated by case C, the
centre of the bead never yields, leaving an elastic core buffered
by plastic collars. Some of the attributes of these three bead structures
can be analyzed in more detail, as we outline in \S \ref{sec:anato}.

\section{Final elasto-plastic anatomy of the beads}\label{sec:anato}

\subsection{Sub-yield-stress deformation}

Where the fluid is locally unyielded, the visco-elastic constitutive
model can be usefully rewritten in terms of
Lagrangian coordinates \cite{renardy95,entov84,clasen06}:
given that $h(0)\approx1$, we
define the ``stretch'' of a material element by $s=h^{-2}$,
and denote time derivatives in the Lagrangian frame by dots.
From (\ref{modeqs}a), it follows that
the axial velocity gradient is $w_z=\dot{s}/s$, and so
the constitutive relations in the elastic regime reduce to
\be
\begin{aligned}
  \mW\left(\dot\tau_{rr} + \frac{\dot{s}}{s}\tau_{rr}\right) &=&
  -\frac{\dot{s}}{s},\\
  \mW\left(\dot\tau_{zz} - 2 \frac{\dot{s}}{s}\tau_{rr}\right) &=&
  2 \frac{\dot{s}}{s}.
\end{aligned}
\ee
Hence,
\be
\begin{aligned}
    \tau_{rr} = \mW^{-1}&(s^{-1}-1),\qquad \tau_{zz} = \mW^{-1}(s^{2}-1),\\
    & \tau_{rr} - \tau_{zz} = \mW^{-1}(h^2 - h^{-4}),\label{elastotaudif}
\end{aligned}
\ee
as long as the local elastic history can be traced back
to the initial moment, when $s(0)\approx1$.
By contrast, if this part of the thread yields at some stage
during its evolution, the fluid there loses
its memory of the initial condition and acquires a certain amount
of unrecoverable plastic strain. In that circumstance,
the final stress difference $\tau_{rr} - \tau_{zz}$ must be determined
from the solution of the initial-value problem. 

\subsection{Final profile equation}

When the bead reaches steady state, (\ref{modeqs}b) reduces to the
force balance equation,
\be
   [h^2 (\mF\{ h \} - \tau_{rr} + \tau_{zz} )]_z = 0
   .
\ee
Because the bead must narrow at the ends to meet the exponentially
thinned string, we may integrate this equation to find that
\be
\mF\{h\} = \tau_{rr}-\tau_{zz} = {\rm Min}
\left[\sqrt{3}\mJ, \mW^{-1}\left( h^2 - h^{-4} \right) \right]
,
\label{eq:finprof}
\ee
as long as \eqref{elastotaudif} applies over the elastic (unyielded) sections.
In view of \eqref{Frel},
the relation in \eqref{eq:finprof} is a second-order ordinary differential
equation (ODE) for the limiting thread profile, $h=h(z)$.
In practice, the center of the bead can be assumed to lie at $z=0$ and
this ODE solved subject to the boundary conditions $h'(0)=0$
and $h\to0$ at the end of the bead, $z=Z$, incorporating the constraint
\begin{equation}
    \int_0^Z h^2 \; {\rm d} z = \half \mL,\label{elastovolcon}
\end{equation}
which follows because the bead contains most of the fluid.

Unfortunately, whenever the final bead is partly unyielded
(as in cases B and C in Fig.~\ref{fig:abc}), this exercise is
confounded by the presence of regions with unrecoverable
plastic strain. Only in the case where the bead becomes fully
plastic ({\it i.e.} case A), is this exercise straightforward.
In particular, in case A, we may solve $\mF\{h\} = \sqrt{3}\mJ$
subject to the constraint \eqref{elastovolcon} and
the boundary conditions $h'(0)=0$ and $h\rightarrow0$ as $z\rightarrow Z$. In the limit $z\rightarrow Z$,
we can also record the local solution,
\be
h \sim \sqrt{ -\frac{4(Z-z)}{\sqrt3 \mJ \log (Z-z)} }
  ,\label{plasprof}
\ee
which deviates from the limiting behaviour for a spherical bead, the expected shape for an Oldroyd-B fluid without a yield stress \cite{clasen06}. 

Fig.~\ref{fig:abc}(a) includes a direct computation
of the final shape from \eqref{eq:finprof}. This profile
matches up with the final shapshot of the numerical
solution of the initial-value problem (and the stress
difference aligns with $\sqrt3\mJ$) except near
the end, where the initial-value computation is still evolving
and visco-elastic relaxation is still taking place
(a feature applying to all three cases in the figure).
Note that the local solution in \eqref{plasprof}
and the profile equation \eqref{eq:finprof} both
highlight how a perfectly plastic bead
does not have constant curvature. Indeed, the
bead in Fig.~\ref{fig:abc}(a) is noticeably
more prolate than the spherical
bead on the string of the visco-elastic problem.

For the other cases in Fig.~\ref{fig:abc}(b,c), we avoid the
regions with unrecoverable plastic strain, and instead
compute the final profile directly from \eqref{eq:finprof}
by shooting from $z=0$, using the final value of $h(0)$ from the
solution of the initial-value problem. This exercise
recovers the shape of the central portion of the bead.
However, on entering the
previously yielded, elastic regions (light blue) in
Fig.~\ref{fig:abc}(b,c), the stress difference departs
significantly from \eqref{elastotaudif}, underscoring our inability
to construct the shape there without adding more information
from the initial-value problem.

\begin{figure}[ht!]
  \centering
  \includegraphics[width=\linewidth]{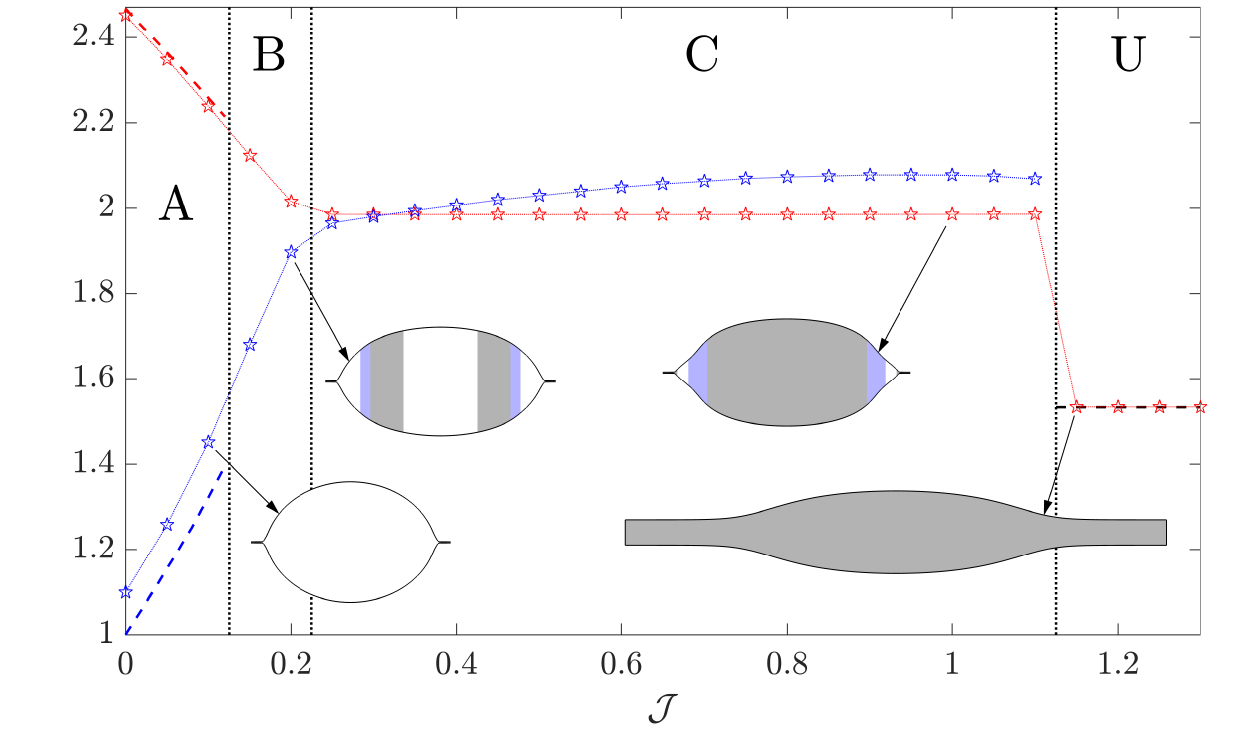}
    \caption{Final maximum radius (red) and aspect ratio (blue) of the beads
      formed for varying $\mJ$, all with
      $\{\mW,\mL,\bet,\mR\}=\{10,20,1,0.01\}$.
      A selection of profiles are also indicated.
      The different bead types are indicated (cases A, B and C;
      see fig \ref{fig:abc}); U refers to the regime wherein
      the thread never yields. Bead aspect ratio is defined as the ratio of
      the half-length of the bead to its maximum radius at the end of a
      simulation. The half-length is defined by approximating the location
      of the bead's edge as the point, $z$, where
      $h(z,t_\mathrm{end})=2\min(h)=0.04$. Dashed lines indicate predictions from solutions to \eqref{eq:finprof} for $h_{\max}$ in the purely plastic case (red), aspect ratio in the purely plastic case (blue), and $h_{\max}$ in the purely elastic case (black). }
    \label{fig:shapej}
\end{figure}

Diagnostics of the
final profiles from a wider suite of computations are shown in
Fig.~\ref{fig:shapej}.
In this suite, the plasto-capillary
number $\mJ$ is varied, holding all the other parameters fixed;
the final maximum radius and bead aspect ratio are presented.
Perhaps counter-intuitively, one finds fully plastic beads (case A)
for the smallest yield stresses, a feature resulting simply
because it is only for the lower values of $\mJ$ that the
bead is able to fully yield.
The fully plastic beads become approximately spherical for $\mJ\to0$,
as indicated by the approach of the aspect ratio toward unity.
Residual visco-elastic relaxation at the bead ends, however,
increases the aspect ratio slightly, as well preventing the
results from matching up with corresponding ones
from directly solving \eqref{eq:finprof} (which are
also included in Fig.~\ref{fig:shapej}).

\begin{figure*}[ht!]
\centering
\begin{subfigure}{.54\textwidth}
  \centering
  \includegraphics[width=\linewidth]{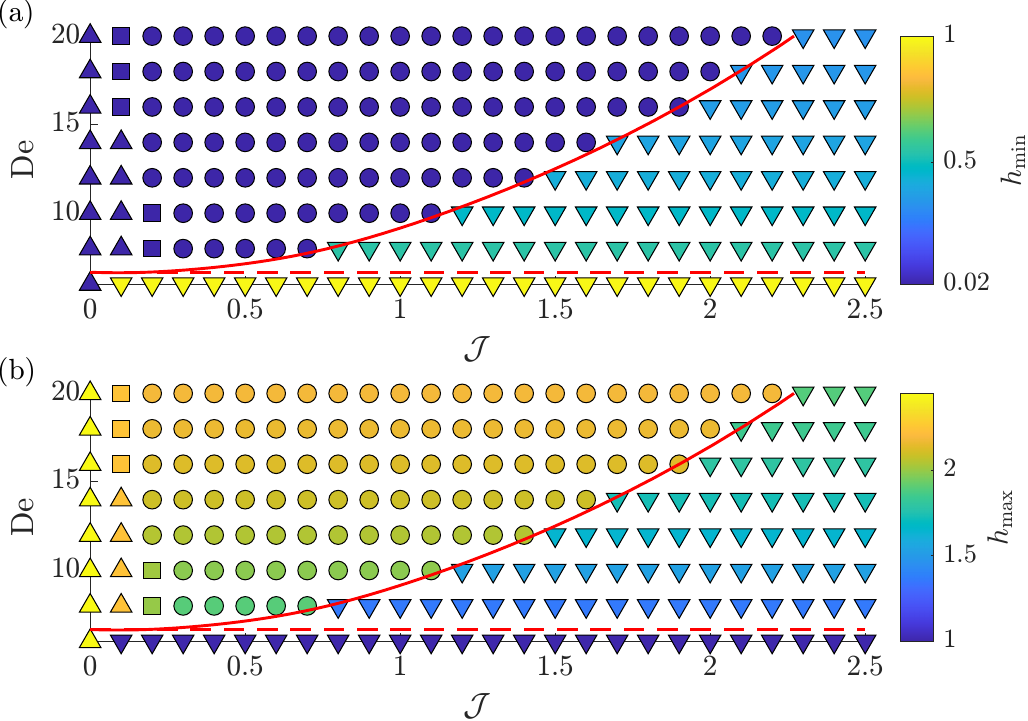}
\end{subfigure}
  \hspace{5pt}
\begin{subfigure}{.44\textwidth}
  \centering
  \includegraphics[width=\linewidth]{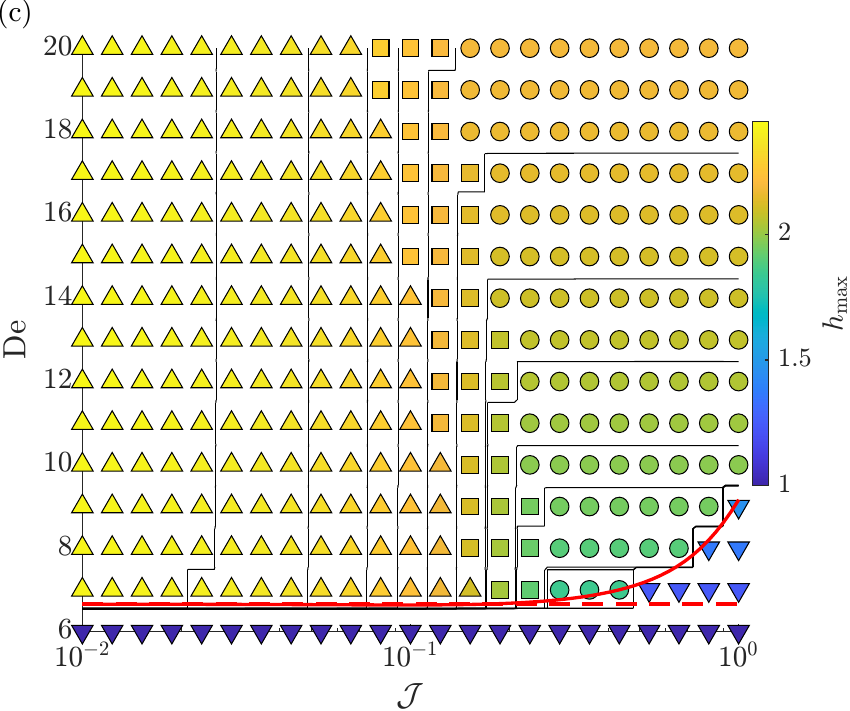}
\end{subfigure}%
\caption{Final (a) minimum and (b,c) maximum radii from computations
      with a range of values of $\mW$ and $\mJ$,
      and $\{\mL,\bet,\mR\} = \{20,1,0.01\}$. 
      The anatomy of the thread is identified by symbol
      (case A, $\blacktriangle$; case B, $\blacksquare$; case C $\bullet$;
      unyielded, $\blacktriangledown$).
      The solid red line shows the locus along which
      $\max_z|\tau_{rr}-\tau_{zz}|=\sqrt{3}\mJ$
      for purely elastic steady states computed from \eqref{eq:finprof}.
      The threshold $\mW=\mW_c=6/(1-k^2)$ for linear instability
      \eqref{LSAcritW} is indicated by the dashed line. The data in (c) correspond to the low-$\mJ$ region of parameter space in (b), but with more data points. The black lines show linearly spaced contours of $h_{\max}$. 
}
\label{fig:carpet}
\end{figure*}

After transitioning to case B, and then to case C, the bead aspect ratio
continues to increase with $\mJ$ in Fig.~\ref{fig:shapej};
all the while the maximum radius decreases. Simultaneously,
the distance between the beads, the length of the strings,
decreases.
Eventually, $\mJ$ reaches the threshold above which the
thread never yields over its thinnest section, prompting a
sharp drop in maximum radius.

At this stage it is worthwile pointing out a key limitation of our results
for the bead structure: our model is based on slender-thread
theory. Following common practice, we have employed the full
form of the curvature in \eqref{Frel} rather than retaining
just the leading term $h^{-1}$. This retention is essential to
furnish a well-behaved model and capture
the spherical shape of the bead in the viscous or visco-elastic
problem. It is, however, a non-asymptotic device that seeks to
add select higher-order terms to the formulation, which otherwise
keeps only the leading order. Any non-slender feature is then
not reliable in the model, including the detailed shape of any
bead whose aspect ratio is order unity.
The elasto-plastic structures that we have identified
are therefore only suggestive of the full richness of 
bead shapes that might arise when the aspect ratio is taken into
account more consistently. In the viscous and
visco-elastic problems, slender-thread theory has been
shown to be qualitatively, if not quantitatively
accurate \cite{eggers08,eggers20}.

\section{Parameter survey}\label{sec:carpetbomb}

We \textcolor{black}{continue} our analysis with a wider survey summarizing the
pinching dynamics over the $(\mJ,\mW)-$plane,
continuing with the settings $\{\mL,\bet,\mR\} = \{20,1,0.01\}$
and the initial conditions in \eqref{ICs}.
For a grid of values of $\mW$ and $\mJ$, we numerically
compute solutions to \eqref{modeqs} and record the final maximum
and minimum radii, along with the character of the anatomy
of the bead. The results are displayed in Fig.~\ref{fig:carpet}.
Our earlier results correspond to following a horizontal line
at $\mW=10$ in this diagram.  

There are two main dividors in Fig.~\ref{fig:carpet}:
first, the linear stability threshold in \eqref{LSAcritW} cuts horizontally
across the  $(\mJ,\mW)-$plane at $\mW\approx6.66$. Below this threshold,
the elastic thread does not suffer Rayleigh-Plateau instability and the
filament relaxes from its initial shape to a new one. The final shape
does not quite correspond to a constant-radius thread: the initial condition
in \eqref{ICs} implies that
\be
\begin{aligned}
    \tau_{rr} &= \mW^{-1}(h^2/h_0^2-1),\qquad \tau_{zz} = \mW^{-1}(h_0^4/h^4-1),\\
     \tau_{rr} &- \tau_{zz} = \mW^{-1}(h^2h_0^{-2} - h_0^4h^{-4})
    \approx 6\mW^{-1}(h - h_0),
\end{aligned}
\ee
where $h_0=1+10^{-3}\cos kz$, rather than \eqref{elastotaudif}.
This leads to a final steady state in which
\be
h \approx 1 + \frac{6\times10^{-3} \cos kz}{6+\mW(k^2-1)} ,
\label{hrelax}
\ee
corresponding to the deformed elastic shape in which capillary
forces become balanced.
Note that the amplitude of the deflection
in \eqref{hrelax} diverges at the linear
stability threshold, indicating that nonlinear terms
are needed to compute properly the final shape nearby.
However, for the parameter values of the grid shown in
Fig.~\ref{fig:carpet}, the final states reached below the
linear stability threshold have deflections that remain small.
Below the dashed line, the final maximum and minimum radii
are therefore close to unity. The one exception is when $\mJ=0$, in which case the thread is an Oldroyd-B fluid. The linear stability threshold \eqref{LSAcritW} does not then apply and linear instability and pinch-off arises for any $\mW$
for the thread length used in the computations (see \S\ref{sec:linstab}).

The second dividor in Fig.~\ref{fig:carpet}
is the yield condition applied to the thinned
section of an unyielded thread that has deformed according to \eqref{elastotaudif}. The stresses become highest at the thinnest point of the thread,
and if the yield threshold becomes breached there, the thread subsequently
continues to pinch-off. In other words, when
\be
\mJ = \frac{1}{\sqrt{3}}
\max_z|\tau_{rr}-\tau_{zz}|\equiv
\frac{1}{\mW\sqrt{3}}\max_z(h^{-4}-h^2)
\label{6.3}
\ee
for the elastic profiles computed from \eqref{eq:finprof},
there is a sudden switch from an elastic final state with
finite minimum radius, to a thread that continually thins toward pinch-off.
The second dividor is shown as the solid line in Fig.~\ref{fig:carpet}.
Below this line, the thread 
behaves as a viscoelastic solid, with $Y=0$ throughout its evolution.
The dynamics is then independent of $\mJ$, as illustrated in
Fig.~\ref{fig:carpet} by the horizontal uniformity of the
results over this part of the parameter plane.

On surpassing the yielding threshold in \eqref{6.3},
the thread forms a beads-on-a-string structure.
For low $\mJ$ (Fig.~\ref{fig:carpet}(c)), each bead becomes fully yielded (case A), and the maximum radius is set entirely by $\mJ$, as indicated by the near-vertical contours of constant $h_{\max}$. At larger $\mJ$, when the bead is mostly unyielded (case C), $\mJ$ instead has almost no impact on the maximum radius;
in Fig.~\ref{fig:carpet}(c)), the contours of constant $h_{\max}$
then becomes nearly horizontal.
The switch in orientation of these contours arises over a
relatively narrow band of intermediate values of $\mJ$, where the bead has the
structure of case B.

\begin{figure}[ht!]
  \centering
  \includegraphics[width=\linewidth]{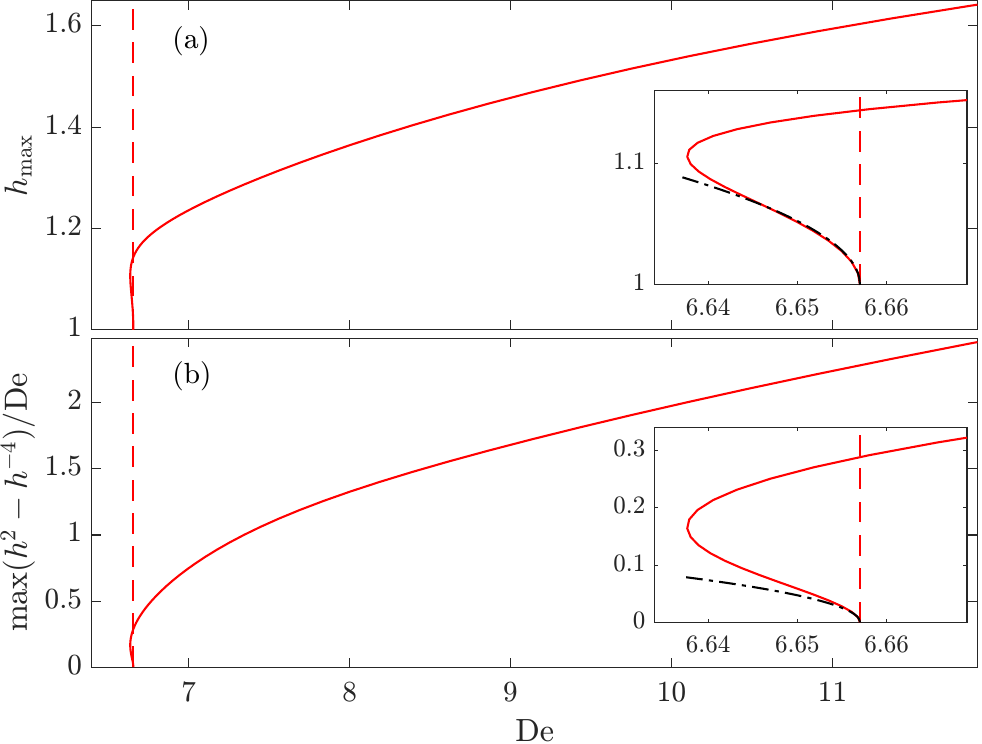}
    \caption{(a) Maximum radius, and (b) maximum stress difference, from purely elastic solutions to the steady-state ODE \eqref{eq:finprof}, with $\mL=20$. Dashed red line indicates the critical Deborah number, $\mW_c$, for linear instability \eqref{LSAcritW}. Dot-dashed black lines in insets show weakly nonlinear approximations, $h_{\max}$ and $6(h_{\max}-1)/\mW$, with $h_{\max}\sim 1+2k(1-k^2)\sqrt{(\mW_c-\mW)/(1+9k^6 + 31k^4 - 5k^2)}$. 
      Note that the curve in (b) and the threshold $W=W_c$,
      after a rotation by $90^\circ$, are
      simply the dividors shown by the solid and dashed lines in
      Fig.~\ref{fig:carpet}.
     }
    \label{fig:bif}
\end{figure}

Note that the yielding threshold \eqref{6.3} does not lie above the linear
stability condition \eqref{LSAcritW} for all values of the yield stress
$\mJ$, because the elastic Rayleigh-Plateau instability is subcritical:
on proceeding along a vertical line in Fig.~\ref{fig:carpet}
past $\mW = \mW_c = 6/(1-k^2)$,
the steady branch of finite-amplitude solutions first bifurcates to
smaller values of $\mW$, as illustrated in Fig.~\ref{fig:bif}.
The branch turns back to higher values of $\mW$ quickly, however, leaving
only a small window of Deborah numbers where multiple solutions
exist and hysteresis is possible.
The subcriticality signifies that there is an abrupt jump in the
final maximum and minimum radius on crossing the horizontal dashed line.
This also means that the maximum stress difference, max$(h^2-h^{-4})/\mW$
(also plotted in Fig.~\ref{fig:bif}) features a similar discontinuity.
Therefore, when that jump exceeds the yield stress $\mJ$, the thinnest
sections of the thread yield immediately and proceed towards pinch-off.
In this circumstance, the yield condition \eqref{6.3} is met by steady elastic solutions that lie below
the linear stability threshold, and the dashed and solid lines in
Fig.~\ref{fig:carpet} must cross. Because the window of values of
$\mW$ is relatively narrow over which there are multiple elastic
solutions, the cross-over of the two dividors is difficult to
observe in Fig.~\ref{fig:carpet}.

\section{Lower solvent viscosity}\label{solv}

Adopting the parameter setting $\bet=1$ ensures that elastic transients
are strongly over-damped during elasto-viscoplastic Rayleigh-Plateau
instability. 
Indeed, a regime diagram like that shown in
Fig.~\ref{fig:carpet} is found for 
$\bet$ ranging
from $0.1$ to values much greater than unity. The only significant
impact of changing the viscosity ratio over this range is that
the relaxation of the beads takes place relatively slowly
for larger $\bet$. As a consequence, for $\bet$ much larger than unity, when the strings thin to the termination criterion
min$(h)=0.02$, the beads remain some way off their final shape,
which obscures any attempt to represent results in the manner
of Fig.~\ref{fig:carpet}.

\begin{figure*}[ht]
\centering
\includegraphics[width=.98\linewidth]{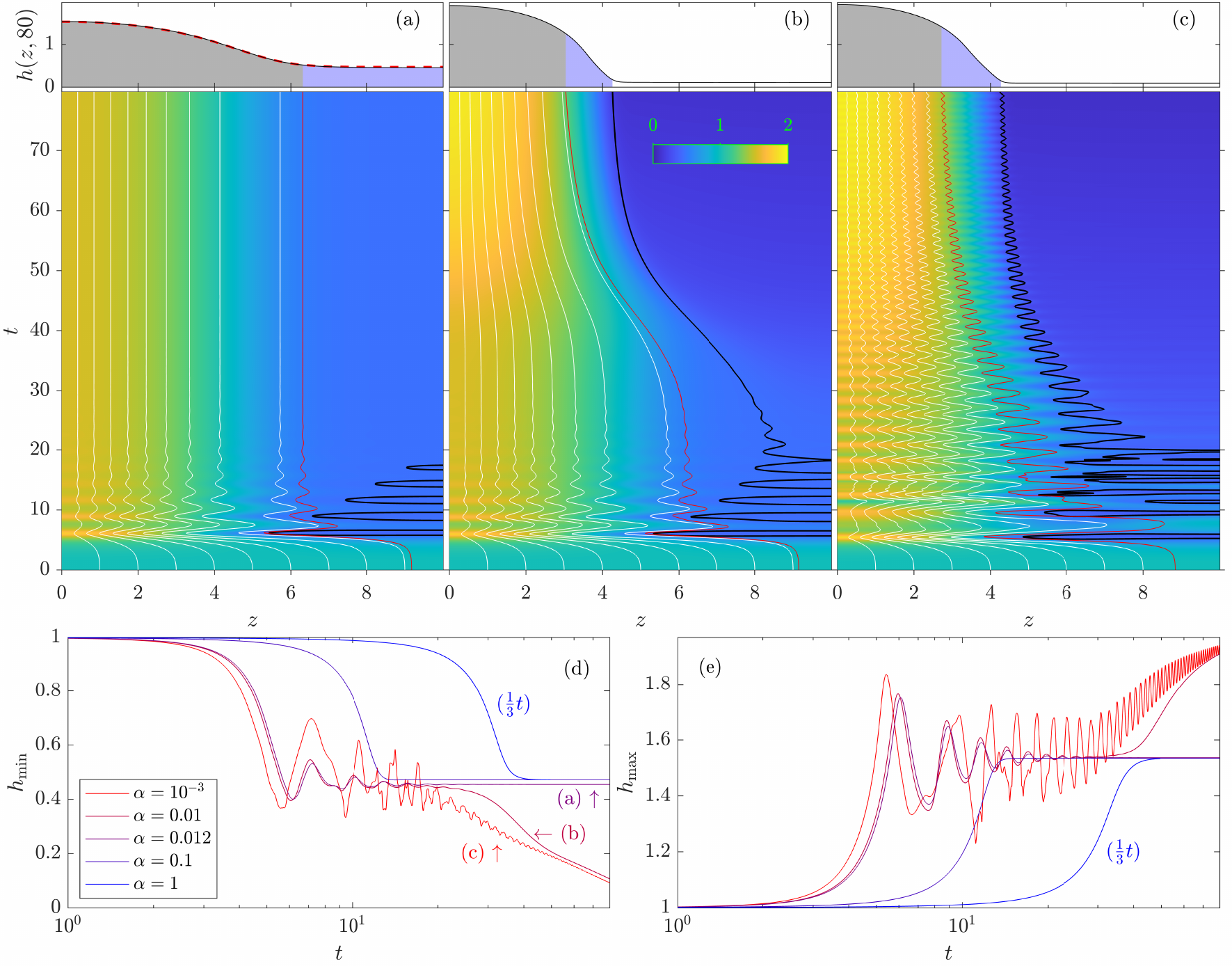}
\caption{
  Numerical solutions for smaller solvent viscosity, $\bet$,
  with $\{\mJ,\mW,\mL,\mR\} = \{1.3,10,20,0.01\}$. 
  (a-c) A snapshot of the thread radius, $h$, at $t=80$ (upper row) above space-time plots of $h(z,t)$, for
  (a) $\bet=0.012$, (b) $\bet=0.01$ and (c) $\bet=10^{-3}$. In the upper panels, as in Figs. \ref{fig:egb}-\ref{fig:shapej}, elastic regions that never
  or previously yield are coloured in grey and blue, respectively;
  the dashed (red) line in (a) shows the final shape for a thread that
  never yields.
  Overlaid on the space-time plots in white and red lines are the paths of sample fluid elements;
  the black lines show the yield surfaces.
  The paths plotted in red border the region that yields at
  some stage during the evolution of the thread.
  \textcolor{black}{Supplementary video 2 shows the evolution of the thread for the example in (b).} 
  Time series of the corresponding minimum and maximum radii
  are shown in panels (d) and (e), along with further solutions
  for other values of $\bet$ (as indicated; the curves are colour-coded, with
  $\bet$ increasing from red to blue). To show the solution
  on the same scale for the time axis, the radii
  for the case with $\bet=1$ are plotted against $t/3$. 
  }
\label{fig:beta}
\end{figure*}

\textcolor{black}{
Qualitatively different dynamics can arise, however, 
by proceeding to much lower values of $\bet$, because
of the possibility that elastic transients then become under-damped
({\it cf.} \cite{lacaze15,moschopoulos23,vpstok,franca2024}).
Although it is unclear whether a low solvent viscosity
in the Saramito model characterizes any real yield-stress
fluid, here we briefly explore whether
elastic oscillations are generated during Rayleigh-Plateau
instability for sufficiently small $\bet$.
Sample solutions with $\bet=0.012$, $0.01$ and $10^{-3}$
for a thread with $\mW=10>\mW_c$ are
illustrated in Fig.~\ref{fig:beta}(a,b,c).}
For all three examples, elastic oscillations appear
as the thread deforms due to linear instability.
\textcolor{black}{
These oscillations are characterized by the sloshing of
fluid back and forth over the thickened
sections of the thread, but become damped out by
finite solvent viscosity. As illustrated by Fig.~\ref{fig:beta},
the damping rate of the oscillations
is reduced as $\bet$ is lowered, but the oscillation frequencies appear
insensitive to this parameter (further computations demonstrate that the
frequency is controlled by the other parameters, $\mW$, $\mR$ and $\mL$).}

\begin{figure*}[ht!]
\centering
\includegraphics[width=.98\linewidth]{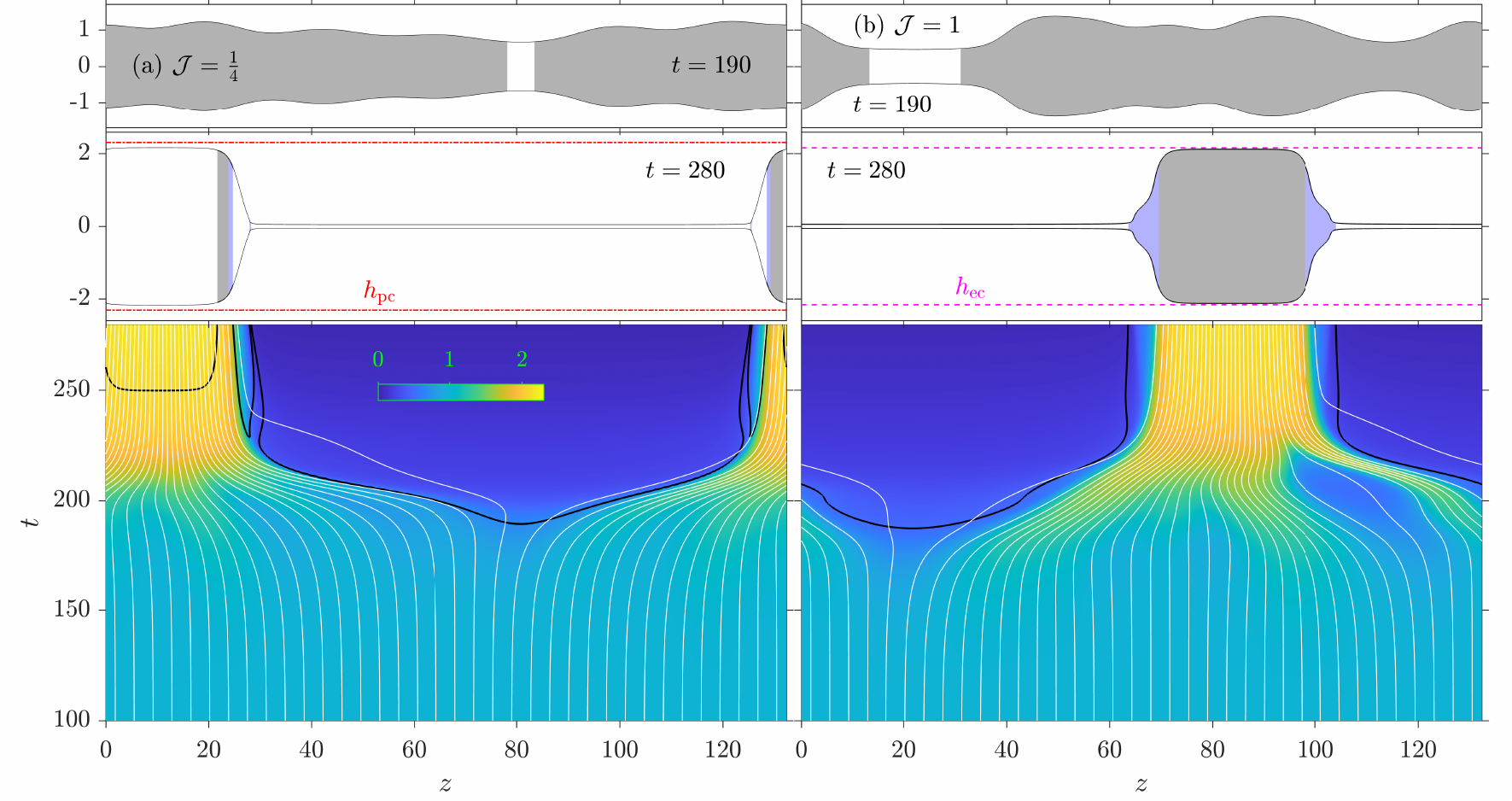}
\caption{\textcolor{black}{Numerical solutions in a longer (periodic) domain, $\mL\approx132$, with $(\mW,\bet,\mR)=(10,1,0.01)$ and (a) $\mJ=0.25$ and (b) $\mJ=1$.
 Initial perturbation is set by \eqref{newIC}.
 The top two panels show the filament's profile at
 $t=190$ and $280$, with shading indicating yielded (white), never yielded
 (grey) and previously yielded (light blue) regions.
 The lower panels present space-time plots, with colour showing $h(z,t)$,
 white lines tracking the paths of individual fluid elements, and thicker
 black lines showing the yield surfaces. In (a), the dot-dashed line
 shows the steady-state
 plastic-cylinder radius, $h_\mathrm{pc}$ in \eqref{hpc}.
 In (b), the dashed line shows the elastic-cylinder radius, $h_\mathrm{ec}$
  in \eqref{hec}.}
  }
\label{fig:long}
\end{figure*}

\textcolor{black}{Although the elastic oscillations are
  eventually damped, their
excitation at finite amplitude and intermediate times can significantly
impact the pinch-off dynamics.}
In particular, for $\mJ=0.13$ and $\mW=10$, the chosen values for the computations in Fig.~\ref{fig:beta}, the yield criterion is not
breached {\it en route} to the final state when $\bet\geq0.1$
(these parameter values lie to the right of
the threshold marked by the red line in Fig.~\ref{fig:carpet}).
For smaller
solvent viscosities, however, transient elastic oscillations
provide a means to locally breach the yield stress.
In the first example of  Fig.~\ref{fig:beta}(a),
yielding only arises during the peaks of the first five elastic
oscillations. As a result,
the thread acquires unrecoverably plastic strain over its
thinnest section, where \textcolor{black}{yielding has} occurred (the blue portion in the upper panel of Fig. \ref{fig:beta}(a)). 
The final minimum radius is then slightly lower
than predicted for a soft visco-elastic solid
as can be seen in Fig.~\ref{fig:beta}(d), which plots
time series of $h_{\min}(t)$ for each of the three solutions
of panels (a,b,c), as well as additional solutions
with large values of $\bet$
(which behave as such solids). The maximum radius
is also slightly higher because of the additional extension
incurred over the thinned section, but this is not obvious in
the time series of $h_{\max}(t)$ shown in
Fig.~\ref{fig:beta}(e).

For the second solution in Fig.~\ref{fig:beta}(b) with $\bet=0.01$, the
yielding of the thread does not subside after the first few oscillations.
Instead, the mean effect eventually
destructively pushes the thinned section towards pinch-off.
Simultaneously,
the bulk of the fluid collects into an elastically deformed bead,
whose periphery acquires unrecoverable plastic strain
(upper panel of Fig. \ref{fig:beta}(b)).
For the final example of Fig.~\ref{fig:beta}(c),
elastic oscillations again trigger pinch-off.
The main difference in this case is that
solvent viscous damping is now \textcolor{black}{sufficiently} small that 
elastic oscillations ring less regularly
with richer frequency content after their initial excitation,
and continue throughout most of the computation
(Fig.~\ref{fig:beta}(e,f)). In other words,
for very low solvent viscosity, elastic oscillations can bypass
the \textcolor{black}{quasi-steady} yield criterion and may
generate persistently ringing bead-on-string structures. 

{
\color{black}

\section{Longer domains}\label{sec:long}

Thus far, we have foscussed on relatively short domains, with $\mL=20$, and
triggered instability by perturbing the radius with the gravest mode,
as in \eqref{ICs}. This ensures the formation of a single bead in the domain,
as may be expected in any domain that is smaller or comparable to the most
unstable wavelength in linear theory \S\ref{sec:linstab}. In much longer
domains, and depending on initial conditions, 
multiple beads may appear and the beads themselves
may become larger to accommodate the increased volume of fluid
in the filament. 

To illustrate the typical dynamics in a longer domain, we compute numerical solutions with $\mL\approx132$, which is twice the most unstable wavelength from linear theory for $(\mW,\bet,\mR)=(10,1,0.01)$. We replace the symmetry conditions in \eqref{latbcs} with periodic boundary conditions at $z=\{0,\mL\}$,
and adopt the initial superposition,
\begin{equation}
  h(z,0) = 1 + \sum_{m=1}^{20}a_m\cos\left(\frac{2m\pi(z-\phi_{m})}{\mL}\right),
  \label{newIC}
\end{equation}
where the amplitudes, $\{a_m\}$, and phase shifts, $\{\phi_m\}$ are chosen randomly, with $|a_m|<10^{-6}$. This setting then allows for competition between growing modes in the initial stages of evolution.
Figure \ref{fig:long} presents two sample numerical solutions.

In both cases, the linear growth phase leads to thinning
in multiple locations, but the yield criterion only becomes
first breached around one of these
(figure \ref{fig:long}, top panels). Yielding then prompts
an accelerated thinning of the filament there. The rest of the filament then
beings to contract into a single bead with a profile
possessing undulations that reflect the earlier elastic deformation.
For the two cases shown in figure \ref{fig:long},
the elastic contraction of the beads eventually 
suppresses those undulations, leading to cylindrical shapes
with rounded ends. For the example in figure \ref{fig:long}(a),
the yield stress is sufficiently low to permit the
centre of the cylindrical bead to yield at late times.

In a small number of other cases that we have computed (not shown), 
the yield criterion becomes breached 
around a second thinned region at intermediate times. The
subsequent contraction away from the two yielded sites then
leads to two cylindrical
beads connected by strings within the domain. However,
for the parameter settings and domain length of figure \ref{fig:long},
the more typical outcome is a single cylinder-on-a-string,
even when the secondary undulations in the profile reach significant
amplitude (as in figure \ref{fig:long}(b) around $(z,t)=(100,190)$).
In other words, the localized yielding 
and accelerated thinning of the filament
inhibit pinch-off elsewhere,
promoting the formation of a single cylinder. In even longer domains,
multiple cylinders likely become more common, although this localized
yielding effect may still reduce their number.


The cylinders that form at late times in figure \ref{fig:long} have relatively 
low aspect ratio in comparison to the beads found for smaller
domains in \S\ref{sec:anato} (implying that our slender-thread theory
is more accurate).
However, the yielding structure within these cylinders is not
too different: the example in figure \ref{fig:long}(a)
is equivalent to a Case B bead, whereas that in
figure \ref{fig:long}(b) corresponds to case C.
Note that, when the beads become cylindrical, their 
final radius follows from simply seeking constant-radius solutions of
\eqref{eq:finprof}: for a fully plastic cylinder, \eqref{eq:finprof}
has solution,
\begin{equation}
  h = h_\mathrm{pc} \equiv \frac{1}{\sqrt{3}\mathcal{J}}.
  \label{hpc}
\end{equation}
For an elastic cylinder, \eqref{eq:finprof} has solution,
\begin{equation}
  h = h_\mathrm{ec} \equiv \left(\sqrt{1+\quar\mW^2}+\half\mW\right)^{1/3}.
  \label{hec}
\end{equation}
Both predictions compare well with the final cylinder radii in
figure \ref{fig:long}, although the solution in (a) requires a
longer computational time to converge closer to \eqref{hpc}. Finally,
as $\mJ$ is decreased, $h_\mathrm{pc}$ increases, demanding that more
fluid is required for a cylinder with this radius to form.
In practice, for very small $\mJ$, large quasi-spherical
beads form instead.

}

\section{Discussion}

A uniform viscoplastic filament cannot suffer the classical Rayleigh-Plateau instability if sub-yield-stress deformation is neglected \cite{dripsI}. Allowing elastic deformation below the yield stress, however, can reintroduce that classical instability. We have investigated the surface-tension-driven instability and nonlinear evolution of an elasto-viscoplastic filament, demonstrating how elastic deformations can give way to yielding and provide a pathway to
eventual pinch-off. 

The unyielded filament behaves as a soft elastic solid, for which the Rayleigh-Plateau instability arises when the Deborah number, $\mW$, exceeds a critical value \cite{matsuo92,barriere,mora2010capillarity,taffetani15a,fu21,pandey21}. The stresses generated during this elastic deformation always become highest over the thinnest sections of the filament. If these stresses exceed the fluid's yield stress, the filament yields first at its thinnest point, then across a longer section where the radius decreases exponentially in time. This provides a pathway to the pinch-off of an elasto-viscoplastic filament from an initially cylindrical configuration due to Rayleigh-Plateau instability. At late times, a beads-on-a-string structure develops, with yielded sections continually thinning while the remaining fluid in the thread contracts into thicker globular structures. 

The structure of the beads is complicated owing to the co-existence of elastic and plastic regions, whose relative proportions are often dictated by the manner in which the bead contracted. In particular, the stress history is needed to properly describe the elastic regions that acquired unrecoverable plastic strain having yielded previously. However, when the plastocapillarity number, $\mJ$, is relatively small, each bead yields completely, and the final shape reflects a balance between capillary forces and the yield stress.
We have quantified 
the structure of the beads, showing how the beads become more elongated as $\mJ$ is increased, with their shape deviating significantly from the near-spherical shape formed in viscoelastic liquid filaments \cite{clasen06}. At sufficiently large $\mJ$, for a given $\mW$, the filament does not yield anywhere and does not
then pinch-off. We have computed the critical $\mW$ for yielding (and pinch-off) as a function of $\mJ$ by computing purely elastic steady states and monitoring
whether their stress fields breach the yielding threshold.
As $\mW$ is increased, sub-yield-stress deformation is enhanced, so pinch-off can occur for increasingly high values of $\mJ$.

In most of our computations, we focused on examples in which solvent
viscosities were relatively high. In this setting, elastic transients
becomes strongly damped and the thread radius evolves monotonically
in time {\it en route} to pinch-off. 
We have, however, also briefly explored
threads with relatively low solvent viscosity, which permits elastic
transients to be under-damped. Decaying elastic oscillations
can then be observed during the Rayleigh-Plateau instability. Although
these oscillations eventually decay for any finite solvent viscosity,
their initial excitation at finite amplitude can force the
thinned sections of the thread to yield at earlier times, which can
open further pathways to pinch-off. 

\textcolor{black}{We also focused on beads-on-a-string formation in a relatively short domain ($\mL=20$). This choice leads to a single bead in the domain,
  permitting us to expose in detail the possible structures and anatomies of individual beads. However, by computing solutions in longer periodic domains,
  we demonstrated how instability can lead to the formation of
  ``cylinder-on-a-string'' structures.
  Pandey et al. \cite{pandey21} have previously explored cylinders-on-a-string
  for solid elastic filaments, concluding that spherical beads
  are more common. 
  For an elasto-viscoplastic filament, however,
  the rapid acceleration of thread
  thinning after yielding can inhibit pinching elsewhere in the filament and
  promote the formation of relatively long cylinders. 
 }

For the Bingham-Oldroyd-B-type elasto-viscoplastic model \cite{saramito07}, extensional stresses prevent pinch-off in finite time. Instead, the thread thins exponentially in time,
similarly to an Oldroyd-B fluid \cite{clasen06}.
Alternative constitutive laws do, however, lead to pinch-off in finite time,
such as those incorporating finite polymer extensibility \cite{renardy95}.
In the elasto-viscoplastic context, Zakeri {\it et al.}~\cite{zakeri25}
have shown that threads pinch off for
the power-law version of Saramito's model \cite{saramito09}.

Finally, we emphasize that
we have employed a slender-thread theory, which provides an appealing reduced-order model that can be interrogated in detail. However, the beads-on-a-string structures formed at late times are not always slender, so these solutions lie outside the range of validity of the asymptotic theory. In our computations, we included the full expression for the free-surface curvature, a commonly-used non-asymptotic extension of the theory which can improve accuracy. \textcolor{black}{However, we await fully three-dimensional axisymmetric simulations to verify the detailed anatomy of the beads-on-a-string structures that we have computed. In a different elongational flow (stretching of an elasto-viscoplastic thread between two plates), Zakeri {\it et al.} \citep{zakeri25} showed that non-negligible shear stresses and two-dimensional flow features can emerge around regions where the free-surface gradients change rapidly. Analogous features may arise for filaments forming beads-on-a-string,
with three-dimensional computations then quantifying the fidelity
of our non-asymptotic modification to slender-thread theory.
}

\section*{Acknowledgements}

JDS was supported by a Leverhulme Trust Study Abroad Studentship. We thank Duncan Hewitt for helpful suggestions. 

\appendix
\section{Stress diffusion}\label{app:numerics}

\begin{figure}
    \centering
    \includegraphics[width=\linewidth]{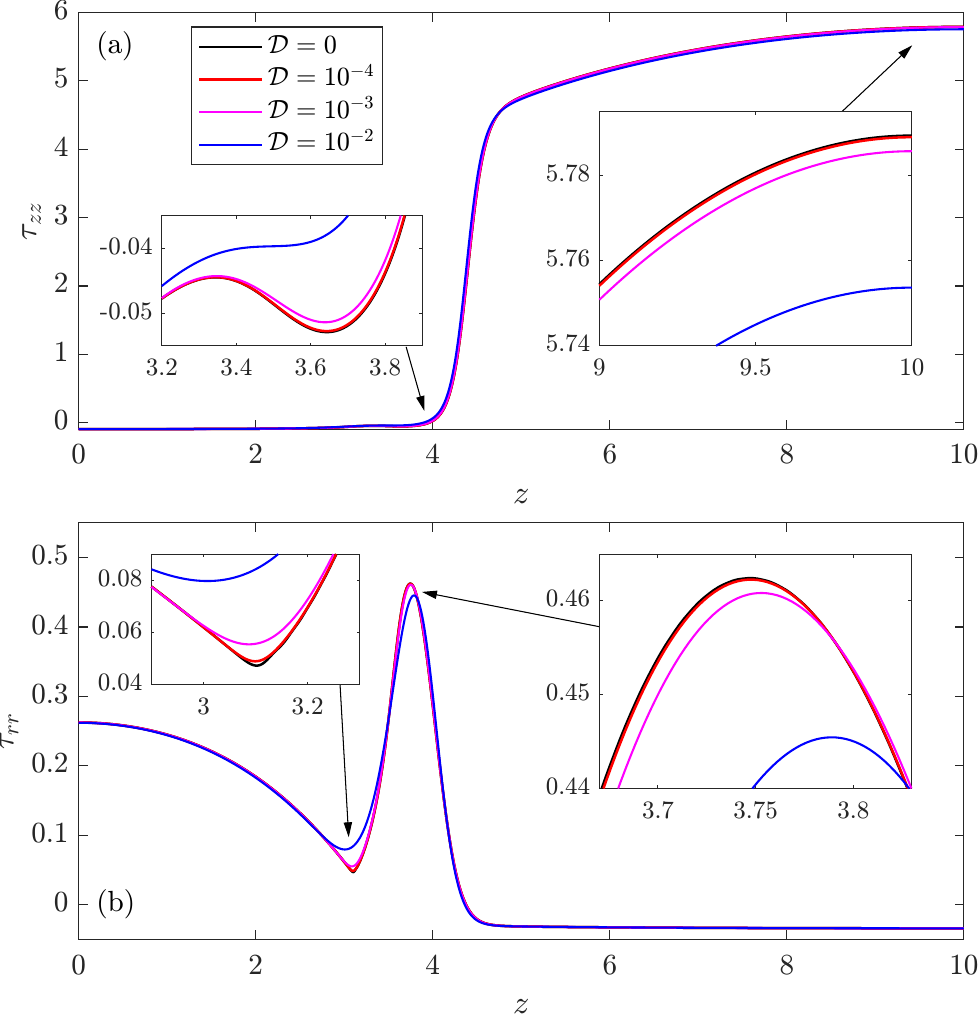}
    \caption{Stress distributions, (a) $\tau_{zz}$ and (b) $\tau_{rr}$, at $t=150$ in simulations with $\{\mW,\mJ,\mL,\bet,\mR\}=\{10,0.4,20,1,0.01\}$ and several values of the stress diffusion coefficient, $\Df$. These parameter values, with $\Df=10^{-4}$, are the same as in Fig \ref{fig:abc}(c). Insets show more detail in areas where the stress diffusion has the most impact.
      }
    \label{fig:diff}
\end{figure}

To ease numerical computations, we introduce stress diffusion, replacing (\ref{modeqs}c,d) with
\begin{equation}
  \mW(\tau_{rr,t} + w\tau_{rr,z} + w_z\tau_{rr}) + Y\tau_{rr} = - w_z + \Df\tau_{rr,zz},
  \label{app:taurr}
\end{equation}
\begin{equation}
  \mW(\tau_{zz,t} + w\tau_{zz,z} - 2w_z\tau_{zz}) + Y\tau_{zz} = 2w_z + \Df\tau_{zz,zz},
  \label{app:tauzz}
\end{equation}
where $\Df\geq0$ is a diffusion coefficient. In numerical computations, we then solve the system of equations (\ref{modeqs}a,\ref{modeqs}b,\ref{app:taurr},\ref{app:tauzz},\ref{Kdefn},\ref{Frel}). 

Fig \ref{fig:diff} illustrates the typical impact of stress diffusion on $\tau_{rr}$ and $\tau_{zz}$ in a sample simulation at a time where significant yielding has occurred and thread thinning is ongoing. The simulations in Fig \ref{fig:diff} have the same parameter values as in Fig \ref{fig:abc}(c), but with varying values of $\Df$. Diffusion typically has its largest impact where stress gradients change rapidly, which occurs around $3\lesssim z\lesssim5$ in Fig \ref{fig:diff}, near the edge of the bead and where the bead connects to the thinning thread (\textit{cf} Fig \ref{fig:abc}c). There is also some effect of diffusion on $\tau_{zz}$ for larger values of $z$; i.e., inside the thinning thread. However, for $\Df=10^{-4}$, the stress distributions are everywhere very close to those from the simulation with $\Df=0$, suggesting that with this value of $\Df$, there is no significant impact on the solution. For the simulations shown in Fig \ref{fig:diff}, the difference in $h_{\max}$ at $t=150$ between $\Df=0$ and $\Df=10^{-4}$ is $3\times10^{-5}$, and the difference in $h_{\min}$ is $5\times10^{-6}$, further verifying that the impact on filament shape is negligible. In all of the simulations presented in the main text, we have used $\Df=10^{-4}$. For $\Df=0$, noticeable numerical error typically begins to appear at late times in simulations. The choice of $\Df=10^{-4}$ ensures accuracy whilst easing computation of solutions at late times.

\bibliographystyle{elsarticle-num}
\bibliography{jfm.bib}

\end{document}